\documentclass[conference]{IEEEtran}
\IEEEoverridecommandlockouts

\usepackage{url}
\usepackage{tikz}
\usepackage{braket}
\usepackage{lipsum}
\usepackage{subfig}
\usepackage{xcolor}
\usepackage{amsmath}
\usepackage{balance}
\usepackage{environ}
\usepackage{physics}
\usepackage{colortbl}
\usepackage{enumitem}
\usepackage{amsfonts}
\usepackage{booktabs}
\usepackage{graphicx}
\usepackage{textcomp}
\usepackage{algorithm}
\usepackage{algpseudocode}
\usepackage{arydshln}
\usepackage[normalem]{ulem}

\newcommand{\sol}{\textit{\text{EnQode}}}

\begin{document}

\title{\sol{}: Fast Amplitude Embedding for Quantum Machine Learning Using Classical Data\vspace{-0.0cm}}

\author{\IEEEauthorblockN{Jason Han}
\IEEEauthorblockA{
\textit{Rice University}\\
Houston, USA}
\and
\IEEEauthorblockN{Nicholas S. DiBrita}
\IEEEauthorblockA{
\textit{Rice University}\\
Houston, USA}
\and
\IEEEauthorblockN{Younghyun Cho}
\IEEEauthorblockA{
\textit{Santa Clara University}\\
Santa Clara, USA}
\and
\IEEEauthorblockN{Hengrui Luo}
\IEEEauthorblockA{
\textit{Rice University}\\
Houston, USA}
\and
\IEEEauthorblockN{Tirthak Patel}
\IEEEauthorblockA{
\textit{Rice University}\\
Houston, USA
}
}

\maketitle

\begin{abstract}

Amplitude embedding (AE) is essential in quantum machine learning (QML) for encoding classical data onto quantum circuits. However, conventional AE methods suffer from deep, variable-length circuits that introduce high output error due to extensive gate usage and variable error rates across samples, resulting in noise-driven inconsistencies that degrade model accuracy. We introduce \sol{}, a fast AE technique based on symbolic representation that addresses these limitations by clustering dataset samples and solving for cluster mean states through a low-depth, machine-specific ansatz. Optimized to reduce physical gates and SWAP operations, \sol{} ensures all samples face consistent, low noise levels by standardizing circuit depth and composition. With over 90\% fidelity in data mapping, \sol{} enables robust, high-performance QML on noisy intermediate-scale quantum (NISQ) devices. Our open-source solution provides a scalable and efficient alternative for integrating classical data with quantum models.

\end{abstract}

\pagestyle{plain}

\section{Introduction}
\label{sec:introduction}

As quantum computing advances toward broader applicability, one of its key challenges is interfacing classical data with quantum algorithms~\cite{wang2022quantumnat,silver2022quilt}. Quantum machine learning (QML) has shown potential in fields ranging from material discovery to the physical sciences, with amplitude embedding (AE) being the critical mechanism for encoding classical data onto quantum states~\cite{diaz2023strongly,garcia2024nuclear,melnikov2023qml}. However, current AE approaches present significant limitations for practical implementation on noisy intermediate-scale quantum (NISQ) devices. Specifically, AE circuits often have a high depth and number of gates, leading to degraded model performance on error-prone architectures, where circuit depth and gate error directly impact output interpretability~\cite{patel2022optic,prasetiadi2024parity}.

Current AE implementations also struggle to maintain a uniform error profile across samples, as the circuit depth and gate composition inherently depend on the data values encoded~\cite{wang2022quantumnas,silver2023mosaiq,cuellar2023time}. The variable depth across samples results in unequal noise exposure -- a problem that can obscure the true performance of quantum algorithms and limit their utility in practical applications. This variability not only exacerbates decoherence issues on quantum hardware but also introduces sample-specific noise, complicating the interpretability and fidelity of downstream QML tasks. \textit{In such settings, the depth and structure of AE circuits need to be managed delicately to balance efficiency with noise resilience~\cite{chen2023quantumsea,silver2023sliq,li2024qust}, underscoring the demand for a new data embedding approach.}

To address these issues, we introduce \sol{}\footnote{\sol{} is published in the Proceedings of the ACM/IEEE Design Automation Conference (DAC), 2025.}, a novel amplitude embedding framework designed to achieve low error and high efficiency through approximate embedding. Rather than attempting exact amplitude representation for each sample, \sol{} leverages a cluster-based approach to approximate AE with high fidelity. Our method begins by grouping data samples using the $k$-means clustering algorithm, enabling us to identify representative states for each cluster that capture the main characteristics of the data distribution. For each cluster, we then train a low-depth, machine-optimized ansatz tailored to the quantum hardware of choice. 

An important aspect of \sol{} is its generalizability across different QML applications with symbolic representation of circuit parameters. Once trained, the cluster-based ansatz allows for the rapid generation of an AE circuit for new data samples by mapping them to their respective clusters and applying transfer learning. This enables \sol{} to yield AE circuits with over 90\% fidelity while maintaining a standardized depth across samples. By reducing the number of physical gates and maintaining a consistent depth across all AE circuits, \sol{} not only enhances fidelity but also ensures uniform error profiles across data samples.

Our evaluation shows that \sol{} achieves significant improvements in efficiency and consistency over the Baseline approach across multiple metrics. Compared to the Baseline, \sol{} reduces the circuit depth by over 28$\times{}$, the single-qubit gate count by over 11$\times{}$, and the two-qubit gate count by over 12$\times{}$, with zero variability across samples due to its fixed ansatz design. In noisy IBM quantum hardware simulation settings, \sol{} demonstrates a state fidelity improvement of over 14$\times{}$ compared to the Baseline, highlighting its robustness to hardware noise. Furthermore, \sol{} reduces the standard deviation of the compilation time over all samples by nearly 3$\times{}$, with an added offline overhead of less than 200 seconds per dataset and class, making it both time-efficient and practical for real-time quantum machine learning applications.
\section{Relevant Background}
\label{sec:background}

\begin{figure*}[t]
    \centering
    \includegraphics[width=0.98\textwidth]{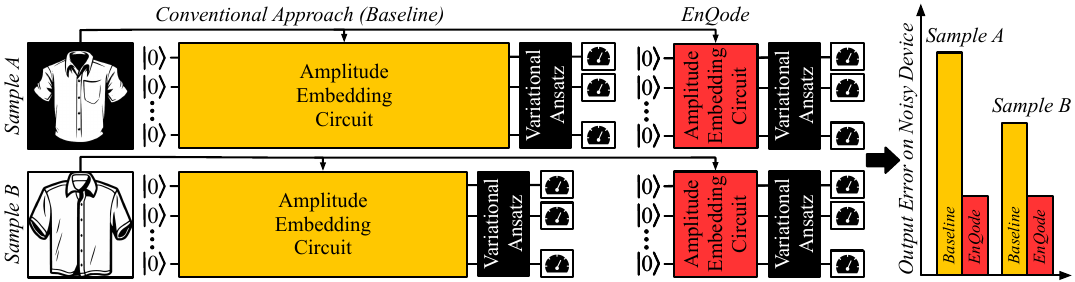}
    \caption{\sol{} provides short-depth and consistent circuits for amplitude embedding across different dataset samples.}
    \label{fig:motivation}
    \vspace{-5mm}
\end{figure*}

\subsection{Qubits, Quantum Gates, and Hardware Noise}

Qubits are the fundamental units of quantum computation, with each qubit in a superposition state $\alpha \ket{0} + \beta \ket{1}$, where $\alpha, \beta \in \mathbb{C}$ and $|\alpha|^2 + |\beta|^2 = 1$. Quantum gates operate on qubits to modify their states, implementing unitary transformations in the form of matrix operations. A unitary operator on $n$ qubits $U \in \mathrm{U}(2^n)$ satisfies the constraints $U^\dagger U = U U^\dagger = I$, where $U^\dagger$ is the conjugate transpose of $U$.

In NISQ devices, gate noise, decoherence, and gate variability significantly impact circuit fidelity, particularly as circuit depth increases~\cite{senapati2023towards,de2022survey,khanal2023evaluating}. The circuit depth $d$ for an amplitude embedding circuit directly correlates with error accumulation, leading to reduced fidelity in the final quantum state $\ket{\psi_{out}}$ due to decoherence effects. Gate error rates also vary across types of gates, with certain gates like SWAP being particularly error-prone -- these gates do not add logical computation to the circuit but help connect distant qubits -- which underscores the need for low-depth and reduced-SWAP implementations for reliable AE~\cite{anagolum2024elivagar,liang2024napa,tomesh2022supermarq}.

\subsection{Variational Quantum Circuits and Optimization}

Variational Quantum Circuits (VQCs) play a central role in QML, leveraging a combination of parameterized gates and classical optimization techniques to minimize an objective function~\cite{de2022survey,li2024quapprox}. A typical VQC can be expressed as:
\begin{equation}
    \ket{\phi(\theta)} = U(\theta) \ket{0}^{\otimes n},
\end{equation}
where $\theta = (\theta_1, \theta_2, \dots, \theta_l)$ are trainable parameters and $U(\theta)$ is a unitary operator constructed from these parameters for an $n$-qubit state. The goal in variational optimization is to find parameters $\theta^*$ that minimize a given loss function $L(\theta)$, often achieved through gradient-based optimization:
\begin{equation}
    \theta^{(t+1)} = \theta^{(t)} - \eta \nabla_\theta L(\theta^{(t)}),
\end{equation}
where $\eta$ is the learning rate.

\subsection{Quantum Data Encoding and Amplitude Embedding}

AE is a technique that encodes classical data into the amplitudes of a quantum state, mapping a classical vector $\vec{f} = (f_1, f_2, \dots, f_{2^n})$ to a quantum state $\ket{\psi}$ such that:
\begin{equation}
    \ket{\psi} = \textstyle\sum_{i=1}^{2^n} f_i \ket{i},
\end{equation}
where $\sum_{i=1}^{2^n} |f_i|^2 = 1$ to satisfy the normalization constraint of quantum states. This process allows classical data to be represented as quantum amplitudes, making it possible to utilize quantum algorithms for ML tasks. In practice, AE requires unitary operators to construct the quantum state $\ket{\psi}$ from an initial state $\ket{0}$~\cite{cuellar2023time,silver2022quilt,patel2022optic}.

\subsection{Challenges with Amplitude Embedding on NISQ Devices}

The reliance on precise amplitude encoding leads to challenges on NISQ devices, where variable-depth circuits introduce inconsistencies in noise exposure across samples. Given that an amplitude embedding circuit for a sample $\vec{f}$ may require $d(\vec{f})$ layers of gates, where $d(\vec{f})$ varies with the encoded values, noise variability becomes an inherent issue:
\begin{equation}
    \ket{\psi(\vec{f})} = U(\vec{f}) \ket{0}^{\otimes n},
\end{equation}
where $U(\vec{f})$ is the unitary operator tailored for encoding $\vec{f}$, and $n$ is the number of qubits required for the embedding. As demonstrated in Fig.~\ref{fig:motivation}, traditional AE circuits lack uniform depth, leading to inconsistent fidelity across samples when used for a downstream variational QML task~\cite{diaz2023strongly,garcia2024nuclear,melnikov2023qml}. 

The effectiveness of VQCs is contingent on the data encoding, which impacts the circuit's ability to represent the problem space accurately. Inconsistent and noisy amplitude embeddings degrade the training process, as each sample faces a different noise profile depending on the circuit depth and composition of its AE circuit. This dependence on high-quality, low-noise encoding underscores the need for efficient AE techniques.

\subsection{Need for \sol{}: Low-Depth Approach to Consistent AE}

\sol{} is designed to overcome the limitations of traditional AE by clustering data samples and representing each cluster with a machine-optimized, low-depth ansatz. By clustering samples into $k$ clusters, \sol{} represents the mean state of each cluster $\vec{c}_i$ using a low-depth ansatz that is optimized offline. For each cluster, we solve for a unitary operator $U(\vec{c}_i)$ that approximates AE with high fidelity:
\begin{equation}
    \ket{\psi(\vec{c}_i)} = U(\vec{c}_i) \ket{0}^{\otimes n}.
\end{equation}

This approach can then be used to perform fast AE for any given dataset sample. Additionally, \sol{} reduces and standardizes the depth across all AE circuits, ensuring that each sample experiences consistent noise exposure, as depicted in Fig.~\ref{fig:motivation}. By achieving high-fidelity, low-noise embeddings, \sol{} provides a robust framework for scalable QML applications, mitigating the issues faced by traditional AE methods on NISQ devices. We describe \sol{} next.
\section{Key Concepts and Design}
\label{sec:design}

We begin by describing the ansatz that \sol{} leverages to amplitude embed any given quantum states from a dataset.

\subsection{\sol{}'s Ansatz for Hardware-Efficient Implementation}

The \sol{} ansatz is designed to leverage the IBM quantum computing hardware basis for optimal efficiency and fidelity (but can be designed for any other hardware basis). On IBM quantum devices, the native gate set consists of the $ECR$ two-qubit gate and the $R_z(\theta)$, $X$, and $SX$ one-qubit gates. Since $R_z$ gates are implemented virtually in software through a frame-of-reference change, they introduce no additional error, making them ideal for parameterized rotations in our ansatz.

\begin{figure}[t]
    \centering
    \includegraphics[width=0.99\columnwidth]{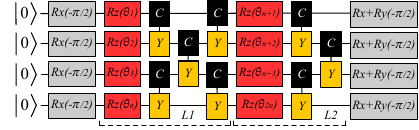}
    \caption{\sol{}'s ansatz design to reduce the circuit depth and the number of physical gates. $CY$ gates are indicated with $C$ for the control qubit and $Y$ for the target qubit.}
    \label{fig:ansatz}
    \vspace{-4mm}
\end{figure}

Because amplitude-embedded states are real-valued with no complex phase, we first apply $R_x(\frac{-\pi}{2})$ gates to each qubit to rotate them from the $\ket{0}$ state in the $z$-$x$ plane to the $\ket{+i}$ state in the $x$-$y$ plane on the Bloch sphere (Fig.~\ref{fig:ansatz}). This setup aligns the qubits in the plane and is suitable for purely $R_z$-based parameterization. Following this, we implement the ansatz shown in Fig.~\ref{fig:ansatz}, which consists of parameterized $R_z$ gates to maintain fidelity, as $R_z$ gates introduce no noise.

For entanglement, we employ controlled-$Y$ ($CY$) gates. Although controlled-$X$, $Y$, and $Z$ gates all require a similar number of $ECR$ and one-qubit gates, thus having similar noise cost, $CY$ gates are preferred here due to their specific interaction properties that preserve the $x$-$y$ plane alignment in the Bloch sphere post-rotation.

The $CY$ gates are arranged in a compact layout that alternates from layer to layer, as depicted in Fig.~\ref{fig:ansatz}, enabling interaction among all qubits within each layer of the ansatz while minimizing gate count. Through extensive experimentation, we identified this alternating configuration as the optimal sparse arrangement on IBM's heavy-hexagonal connectivity layout (linear section). It requires no additional SWAP gates, maximizing fidelity and minimizing error per layer.

We apply multiple layers of this structured ansatz, with each layer refining the AE while ensuring consistent qubit interactions. Finally, $R_x(-\frac{\pi}{2})$ and $R_y(-\frac{\pi}{2})$ gates are applied to return the qubits to the $z$-$x$ plane, completing the embedding process with minimal depth and maximal fidelity. This hardware-efficient design directly reduces error-prone physical gates and provides a high-fidelity, low-noise AE circuit.

\subsection{Symbolic Representation for Efficient Optimization}
\label{subsec:symb_rep}

\begin{figure}[t]
    \centering
    \includegraphics[width=0.98\columnwidth]{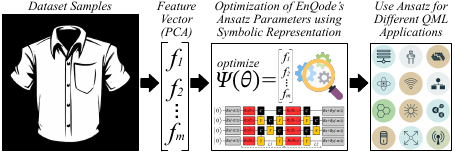}
    \caption{\sol{} employs symbolic representation for fast optimization of its amplitude embedding circuit ansatz.}
    \label{fig:symbolic}
    \vspace{-4mm}
\end{figure}

A key innovation in \sol{} is its use of symbolic representation to achieve fast and accurate amplitude embedding. Unlike traditional approaches that rely on numerical approximations for each optimization step, \sol{} represents the parameters of all $R_z$ gates symbolically, allowing the amplitude embedded state vector $\ket{\psi(\theta)}$ to be fully expressed as an analytical function of these parameters. Given $n$ qubits, let $\theta = (\theta_1, \theta_2, \dots, \theta_l)$ denote the set of $R_z$ gate parameters in the ansatz. The intermediate state vector $\ket{\psi(\theta)}$ that is optimized in the ansatz can be written as:
\begin{equation}
\ket{\psi(\theta)} = \frac{1}{\sqrt{2^n}}\begin{bmatrix}
    k_1 e^{i\frac{\sum_{j=1}^l p_{1j} \cdot \theta_j}{2}} \\
    k_2 e^{i\frac{\sum_{j=1}^l p_{2j} \cdot \theta_j}{2}} \\
    \vdots \\
    k_{2^n} e^{i\frac{\sum_{j=1}^l p_{2^nj} \cdot \theta_j}{2}}
\end{bmatrix},
\end{equation}
where \(p_{rj} \in \{ -1, 0, 1 \} \) and \(k_r \in \{  -1, -i, i, 1  \}\), for all \( 1 \leq r \leq 2^n, 1 \leq j \leq l \).

We make this symbolic representation to leverage the following insight: by converting the amplitudes into relative phases, we are able to have a compact expression with explicit, simple partial derivatives (as we simply take the derivative of an exponential function composed with a linear function), which we can leverage in gradient-based optimization. After optimizing the fidelity for our statevector in this compact representation, we perform a change of basis to convert the relative phases back to real-valued amplitudes used for amplitude embedding. 

This approach allows for direct calculation of the impact of each $R_z$ parameter on the state vector, providing an efficient and precise mapping from the circuit’s symbolic parameters to the target amplitude values. This symbolic linkage between parameters and output state is unique in that it eliminates the need for repeated numerical recalculations, which are computationally expensive and prone to rounding errors in conventional approaches.

The optimization task in \sol{} is to match each element $\psi_i(\theta)$ of the parameterized state vector to the target real amplitude value $x_i$, representing the cluster mean. Specifically, \sol{} seeks the optimal parameters $\theta^*$ such that:
\[
\psi_i(\theta^*) \approx x_i \quad \text{for each } i,
\]
where $x_i$ are the target values from the data vector.

The symbolic representation with simple derivatives enables \sol{} to leverage an efficient gradient-based optimization algorithm. For this work, we use the Limited-memory BFGS algorithm because it empirically performed the best compared to other algorithms we tested~\cite{2020SciPy-NMeth}~\cite{liu1989limited}. As a quasi-Newton method, it relies on accurate Jacobian computations, which are computed efficiently through our symbolic representation. We compute gradients and estimate the inverse Hessian by supplying a symbolic representation of the Jacobian, allowing for fast gradient evaluation.

As visually represented in Fig.~\ref{fig:symbolic}, this approach dramatically accelerates convergence, allowing \sol{} to complete training faster than traditional iterative methods that require extensive numerical calculations and lack direct control over each parameter’s effect on the final state. This unique design forms the foundation of \sol{}’s ability to achieve rapid, high-fidelity amplitude embeddings.

\subsection{\sol{}'s Use of Clustering for Offline Training}

To enable efficient and noise-resilient AE, \sol{} uses $k$-means clustering~\cite{likas2003global} to partition the dataset into $k$ clusters. Each sample $\vec{x}$ in the dataset is assigned to one of these clusters based on its similarity to the cluster centroids, producing $k$ clusters of similar data samples. For each cluster, we calculate the mean sample, $\vec{c}_i$, representing the central characteristics of that cluster. This mean sample serves is used to train the embedding circuit for each cluster.

Once the clusters are defined, \sol{} performs ``offline'' training by applying the hardware-optimized ansatz discussed previously to the mean sample of each cluster. Using the symbolic representation optimization technique, the ansatz parameters for each cluster’s mean sample $\vec{c}_i$ are optimized to maximize fidelity with minimal circuit depth. As this training approach leverages symbolic representations, it significantly reduces the computational overhead, completing the training for all clusters for a given class in a dataset in under 200 seconds. This rapid optimization allows each cluster’s ansatz to be fine-tuned to capture the structure of its mean sample, effectively encoding the main features of all samples within the cluster.

\begin{figure}[t]
    \centering
    \includegraphics[width=0.98\columnwidth]{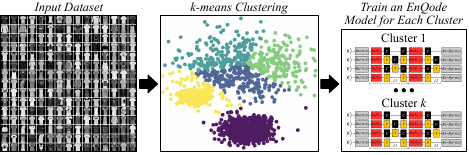}
    \caption{\sol{} divides a dataset into multiple clusters and trains a symbolic representation model for each cluster.}
    \label{fig:clustering}
    \vspace{-5mm}
\end{figure}

As shown in Fig.~\ref{fig:clustering}, this clustering-based approach allows \sol{} to achieve high-fidelity amplitude embedding efficiently across diverse datasets. By training on cluster means, \sol{} reduces redundant computations and ensures that the embedding circuit depth and noise profile remain consistent, thereby providing a scalable solution for QML applications.

The trained cluster models are then stored and used to support ``online'' training and inference, as we describe next.

\subsection{Transfer Learning for Fast Amplitude Embedding}

With thousands or millions of samples in a dataset, it is critical that these states are able to be amplitude embedded into quantum circuits as quickly as possible (in ``real-time'') for fast training and inference. \sol{} uses transfer learning from pre-trained cluster models to enable fast amplitude embedding for incoming samples. When a new sample $\vec{x}_{\text{new}}$ arrives, it is assigned to the nearest cluster based on its Euclidean distance to the cluster centroids. This quick assignment identifies the most relevant pre-trained model for the sample, allowing \sol{} to initialize $\vec{x}_{\text{new}}$'s embedding circuit with the optimized parameters of the corresponding cluster mean. The parameters are then fine-tuned from that initialization using the same symbolic optimization algorithm described in Section~\ref{subsec:symb_rep}.

\begin{figure}[t]
    \centering
    \includegraphics[width=0.98\columnwidth]{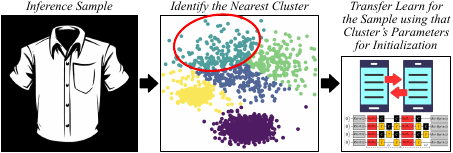}
    \caption{An inference sample gets matched to one of the clusters, and its amplitude embedding circuit is quickly trained using the cluster's trained model to initialize the parameters.}
    \label{fig:transfer}
    \vspace{-5mm}
\end{figure}

As shown in Fig.~\ref{fig:transfer}, this initialization provides a close approximation for the embedding of $\vec{x}_{\text{new}}$, minimizing the need for further optimization and significantly accelerating the embedding process. By beginning with cluster-trained parameters, \sol{} achieves high-fidelity AE with consistent circuit depth and noise profile across samples, all without retraining from scratch. This transfer learning approach ensures that \sol{} can embed both training and inference samples rapidly, making it well-suited for large datasets and real-time quantum machine learning workflows.

\subsection{Putting All of \sol{}'s Design Elements Together}

\sol{} integrates clustering, symbolic optimization, and transfer learning to create an efficient AE framework. By clustering data and training a hardware-optimized ansatz for each cluster mean using symbolic representation, \sol{} achieves a rapid, low-overhead setup with optimized parameters. For each new sample, transfer learning initializes its embedding circuit using the pre-trained parameters of the nearest cluster, enabling consistent, fast embedding without extensive re-optimization.

The ansatz design maximizes hardware efficiency through virtual $R_z$ gates and a sparse $CY$ layout, minimizing circuit depth and noise. \textbf{Together, these elements provide \sol{} with a scalable, high-fidelity solution suitable for real-time quantum machine learning on NISQ devices, maintaining uniform performance across samples.}
\section{Methodology}
\label{sec:methodology}

\subsection{Experimental Setup and Framework}

We run our compliation experiments and classical processing tasks on our local computing cluster. The cluster consists of nodes with the AMD EPYC 7702P 64-core processor with x86\_64 architecture and a 2.0 GHz clock. We spawn virtual machines (VMs) on these nodes consisting of 8 cores, 32 GB memory, and 32 GB storage for each of our experiments, providing more than sufficient resources for each experiment. The VMs are resource-bounded and not overprovisioned, ensuring that each experiment has exclusive access to the hardware resources assigned to it without any interference, which helps us provide accurate and consistent timing analysis.


For implementing and evaluating \sol{}, we wrote Python modules. We used 
\texttt{qiskit}~\cite{aleksandrowiczqiskit} for representations of symbolic and parameterized circuits, as well as for density matrix fidelity calculations. We used \texttt{qiskit\_aer} to perform ideal and noisy density matrix simulations given the characteristics of the \texttt{ibm\_brisbane} quantum computer~\cite{castelvecchi2017ibm}. We used \texttt{scipy}~\cite{2020SciPy-NMeth} to implement the L-BFGS optimization algorithm. For all experiments, we used 8-qubit circuits, with an 8-layer ansatz structure as depicted in Fig.~\ref{fig:ansatz}. The number of clusters is chosen such that the \textit{state fidelity} (defined below) between any datapoint and its nearest cluster is at least 0.95.

For implementing and evaluating the Baseline, we use the \texttt{qiskit} state preparation algorithm~\cite{aleksandrowiczqiskit}~\cite{iten2016quantum}. For transpilation onto IBM hardware, we use the transpilation optimization level of 0 to eliminate the influence of confounding factors such as synthesis techniques for our analysis. All of our metrics are evaluated using circuits transpiled to \texttt{ibm\_brisbane}.

\subsection{Evaluation Datasets}

To evaluate \sol{}, we use three standard image datasets: MNIST~\cite{mnist}, Fashion-MNIST (F-MNIST)~\cite{xiao2017fashion}, and CIFAR-10 (CIFAR)~\cite{krizhevsky2009learning}. We randomly sampled 5 classes for each dataset, and for each class, we randomly sampled 500 images from that class. For each dataset, we apply Principal Component Analysis (PCA)~\cite{mackiewicz1993principal} to reduce dimensionality, focusing on retaining the most significant features. The resulting features are then normalized to ensure compatibility with AE. This preprocessing step allows for efficient, high-fidelity embedding of each dataset in \sol{}'s framework. 

\subsection{Assessment Metrics}

We compare \sol{}'s performance to the conventional exact amplitude embedding technique~\cite{schuld2019quantum}~\cite{iten2016quantum} -- referred to as Baseline. For all circuit-based metrics, we exclude $R_z$ gate counts, as these can be implemented virtually. First, we assess the \textit{circuit depth} and the \textit{total number of gates} used for amplitude embedding with Baseline and \sol{} -- we examine the mean and standard deviation behaviors. These are the key metrics that affect circuit fidelity on NISQ devices. We also look at the \textit{number of physical one-qubit and two-qubit gates}. Second, we evaluate the \textit{compilation times} to generate the amplitude embedded circuits. We also investigate the \textit{state fidelity} of how closely the amplitude embedded state matches the desired state. We use a common metric for mixed states defined as \(F(\rho, \sigma) = \left( tr \sqrt{\sqrt{\rho}\sigma\sqrt{\rho}} \right)^2\)~\cite{jozsa1994fidelity}, where $\rho$ and $\sigma$ are the two quantum states (one desired and one measured). We perform these assessments in ideal and noisy simulation scenarios.
\section{Evaluation and Analysis}
\label{sec:evaluation}

\begin{figure}[t]
    \centering
    \includegraphics[width=0.485\columnwidth]{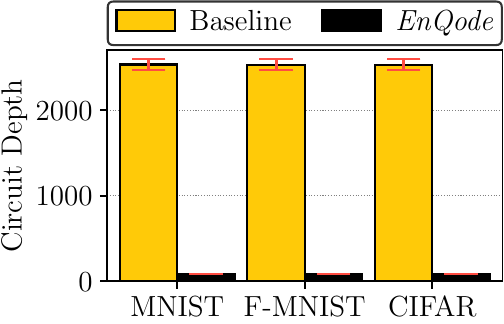}
    \hfill
    \includegraphics[width=0.485\columnwidth]{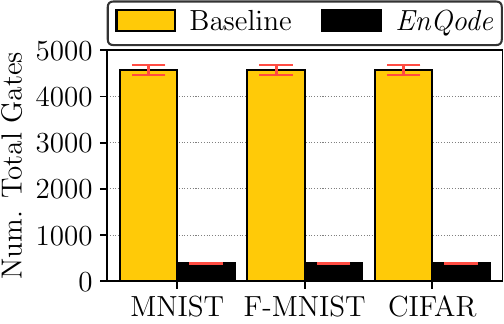}
    \vspace{1mm}
    \hrule
    \vspace{1mm}
    \caption{\sol{} achieves orders of magnitude lower circuit depth and total number of gates than the Baseline, with no variability due to fixed circuit -- the Baseline displays variability for both metrics.}
    \vspace{-5mm}
    \label{fig:depth}
\end{figure}

\begin{figure}[t]
    \centering
    \includegraphics[width=0.51\columnwidth]{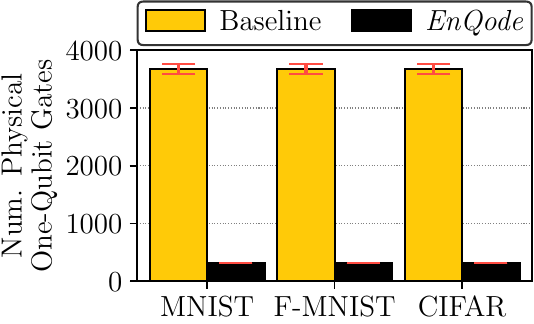}
    \hfill
    \includegraphics[width=0.46\columnwidth]{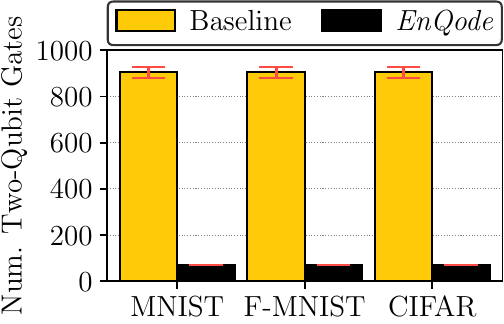}
    \vspace{1mm}
    \hrule
    \vspace{1mm}
    \caption{\sol{} also achieves orders of magnitude less one-qubit and two-qubit physical gates than the Baseline and has no variability due to a fixed circuit design -- in contrast, the Baseline displays variability for both metrics.}
    \vspace{-5mm}
    \label{fig:gates}
\end{figure}

\subsection{\sol{} Reduces the Circuit Depth and Two-Qubit Gate Count (and Their Variability) Over the Baseline Approach}

As shown in Fig.~\ref{fig:depth}, \sol{} achieves a significant reduction in circuit depth (28$\times{}$) on average and total gate count (12$\times{}$) on average compared to the Baseline, with the added advantage of complete consistency across data samples. The Baseline approach suffers from variability in circuit length due to the data-dependent nature of traditional AE, resulting in fluctuations in both circuit depth and the number of gates required. In contrast, \sol{} standardizes the ansatz length and gate composition, ensuring a fixed circuit structure for each dataset.


Similarly, Fig.~\ref{fig:gates} illustrates that \sol{} significantly reduces both one-qubit (11$\times{}$) and two-qubit (12$\times{}$) gate counts compared to the Baseline while completely eliminating gate count variability. By using a fixed arrangement of one-qubit $R_z$ gates and a sparse $CY$ entangling structure, \sol{} ensures that each sample faces a uniform noise profile, essential for stable performance in QML applications. The Baseline approach shows high variability in both gate types, with some samples requiring up to 100 more gates than others (in contrast, the total number of gates used in the ansatz in EnQode is 384).

The reduction in circuit depth, single-qubit gate count, 
two-qubit gate count, and variability provided by \sol{} directly enhances reliability in noisy quantum computing environments. With a fixed circuit structure, \sol{} minimizes exposure to noise, ensuring that each sample is treated uniformly. This consistency is particularly beneficial on NISQ devices, where circuit variability directly impacts fidelity. 

The stability and reduced complexity of \sol{}'s circuits are expected to yield improved performance under noise, a hypothesis we test next.

\begin{figure}[t]
    \centering
    \subfloat[Ideal Simulation]{\includegraphics[width=0.485\columnwidth]{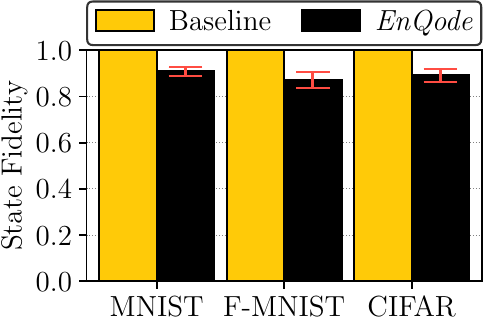}}
    \hfill
    \subfloat[Noisy Simulation]{\includegraphics[width=0.485\columnwidth]{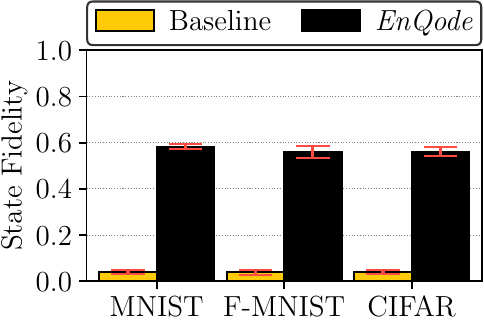}}
    \vspace{1mm}
    \hrule
    \vspace{1mm}
    \caption{(a) \sol{} achieves high state fidelity in ideal simulation. (b) In addition, it also achieves a higher fidelity than the Baseline in noisy NISQ simulation settings.}
    \vspace{-5mm}
    \label{fig:fidelity}
\end{figure}


\subsection{\sol{}'s Generated Circuits Achieve High and Consistent State Fidelity in the face of Hardware Noise Effects}

To assess \sol{}'s effectiveness in representing a given data point state, we evaluate state fidelity across ideal simulations and noisy hardware simulation executions.

In ideal simulations (Fig.~\ref{fig:fidelity}(a)), \sol{} achieves an average state fidelity of 89\% when transpiled to real hardware (where Baseline achieves a state fidelity of 100\%, as it performs exact amplitude embedding)~\cite{iten2016quantum}. Although \sol{} has worse fidelity compared to Baseline in ideal simulation, we show that, on average, we can embed a state that is quite similar to the target state of interest, while significantly reducing both circuit depth and gate count, as discussed above.

In addition, under noisy simulation conditions (Fig.~\ref{fig:fidelity}(b)), \sol{} significantly outperforms the Baseline (by over 14$\times{}$). The Baseline circuits, which rely on deep circuits with variable lengths, suffer from high error accumulation in noisy environments, leading to substantial variability in state fidelity across samples. In contrast, \sol{} maintains higher fidelity and reduced variability due to its consistent, low-depth circuit design. This stability in noisy settings underscores the advantage of \sol{}’s fixed-length ansatz, which minimizes and standardizes exposure to noise-induced errors.


Overall, \sol{}’s ability to maintain higher fidelity in noisy hardware environments demonstrates its practicality for quantum machine learning on NISQ devices. By ensuring lower circuit depth and consistent gate structure, \sol{} provides a reliable amplitude embedding approach that is more robust to noise, a crucial attribute for achieving stable performance in real-world quantum applications.

\begin{figure}[t]
    \centering
    \subfloat[Compilation Times]{\includegraphics[width=0.515\columnwidth]{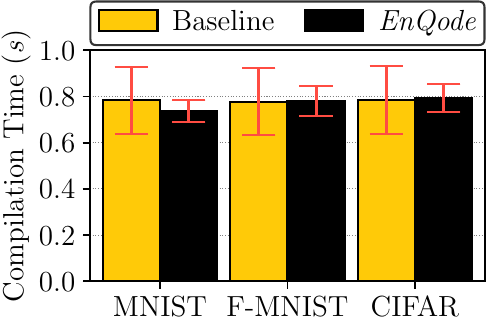}}
    \hfill
    \subfloat[\sol{}'s Time Overhead]{\includegraphics[width=0.45\columnwidth]{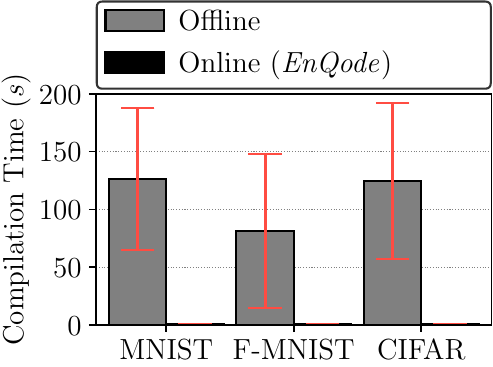}}
    \vspace{1mm}
    \hrule
    \vspace{1mm}
    \caption{(a) \sol{} achieves lower compilation times than the Baseline. (b) \sol{}'s online compilation time is low. Its offline compilation time (for training the clusters) is also low.}
    \vspace{-5mm}
    \label{fig:times}
\end{figure}

\subsection{\sol{} Achieves Lower Compilation Time Compared To the Baseline and has a Low Offline Compilation Time}

To evaluate \sol{}'s efficiency in generating AE circuits, we compare its compilation times with the Baseline, breaking down the online and offline components. Fig.~\ref{fig:times}(a) shows that \sol{} achieves nearly identical compilation time with almost 3$\times{}$ reduced standard deviation compared to the Baseline due to \sol{}'s symbolic representation and transfer learning from pre-trained cluster means. Since \sol{}'s online compilation starts optimization using parameters from the closest cluster mean, it remains highly efficient for embedding new samples in real-time with low variability in compilation time.

In Fig.~\ref{fig:times}(b), we provide a comparison of \sol{}'s offline and online compilation times. The offline compilation step involves training the ansatz for each cluster mean, which is performed only once per dataset and class and thus introduces a minimal overhead of less than 200 seconds. 

As a result, \sol{} achieves rapid online compilation without sacrificing fidelity or adding significant preprocessing costs. The combination of low offline and online compilation times underscores \sol{}'s practicality for real-time QML applications, where speed and efficiency are crucial.

\section{Related Work}
\label{sec:related_work}

The challenge of encoding classical data onto quantum circuits has spurred numerous methods for state preparation and feature embedding, but none directly address the unique requirements of AE with noise resilience and consistency across samples. Existing approaches primarily focus on general state preparation techniques~\cite{gleinig2021efficient, zylberman2024efficient, melnikov2023quantum, araujo2021divide, stetcu2023projection, wang2024quantum}, which can lead to variable-depth circuits that suffer from inconsistencies in noise exposure and fidelity.

Quantum transfer learning~\cite{otgonbaatar2023quantum} offers valuable insights into high-dimensional data adaptation but does not cater to AE’s demand for low-noise, standardized circuit structures. Similarly, applications in drug discovery~\cite{li2021drug} and image classification~\cite{alam2021quantum} integrate quantum embedding methods. Yet, they rely on embedding mechanisms that often lack circuit depth control, making them unsuitable for AE, where consistent noise handling is critical for uniform sample treatment.

More recent efforts have tackled feature-specific embeddings~\cite{thumwanit2021trainable} and variational techniques~\cite{chakrabarti2021iccad}, optimizing for fidelity and efficiency in QML. However, these methods do not account for AE-specific needs like fixed circuit depth and low-noise environments for high-fidelity encoding. \sol{} introduces the first AE-specific approach that achieves consistent, high-fidelity embeddings by clustering data samples and designing a low-depth, machine-optimized ansatz. By standardizing circuit depth and minimizing gate variability across samples, \sol{} ensures uniform noise resilience, setting a new standard for practical AE in scalable QML.
\section{Conclusion}
\label{sec:conclusion}

In this work, we introduced \sol{}, an efficient AE framework designed to address the limitations of the traditional AE method on quantum devices. By employing a cluster-based approach with low-depth, symbolic representation ansatz, \sol{} achieves consistent circuit depth and gate composition across samples, effectively mitigating noise variability and high error rates. Our solution delivers over 90\% fidelity in the fast embedding of classical data, providing a robust foundation for QML. As an open-source tool, \sol{} offers a practical pathway for more reliable data encoding, making it well-suited for diverse applications on quantum platforms.

\vspace{2mm}

\noindent\sol{}'s code is open-sourced at: {\small\texttt{\url{https://github.com/positivetechnologylab/EnQode}}}.
\section*{Acknowledgment}

This work was supported by Rice University, the Rice University George R. Brown School of Engineering and Computing, and the Rice University Department of Computer Science. This work was supported by the DOE Quantum Testbed Finder Award DE-SC0024301 and the National Science Foundation Award NSF-DMS 2412403. This work was also supported by the Ken Kennedy Institute and Rice Quantum Initiative, which is part of the Smalley-Curl Institute.

\balance


\bibliographystyle{plain}
\bibliography{main}

\end{document}